\documentclass[aps, pra, reprint, amsmath, amssymb, longbibliography, superscriptaddress]{revtex4-1}

\usepackage[T1]{fontenc}
\usepackage{palatino}
\usepackage{soul}
\usepackage{lipsum}
\soulregister\ref{7}
\usepackage{graphicx}
\usepackage[svgnames]{xcolor}
\usepackage{amsmath, bm}
\usepackage[detect-none]{siunitx}
\usepackage{fancyref}
\usepackage[colorlinks,
            linkcolor=Blue,
            urlcolor=Blue,
            citecolor=DarkGreen]{hyperref}
\usepackage[utf8]{inputenc}
\usepackage{textcomp}
\usepackage[version=3]{mhchem}
\usepackage{miller}
\usepackage{palatino} 

\usepackage{pdfpages}
\usepackage{pgffor}
\usepackage{etoolbox}
\makeatletter
\patchcmd{\@outputpage@head}{\@ifx{\LS@rot\@undefined}{}{\LS@rot}}{}{}{}
\makeatother

\begin{document}

\title[]{Large tunability of strain in \ce{WO3} single-crystal microresonators\\controlled by exposure to \ce{H2} gas}

\author{Nicola~\surname{Manca}$^{\dag}$}
\email{manca@fisica.unige.it}
\affiliation{Kavli Institute of Nanoscience, Delft University of Technology, P.O. Box 5046, 2600 GA Delft, The Netherlands}
\affiliation{Dipartimento di Fisica, Università degli Studi di Genova, via Dodecaneso 33, Genova, Italy}
\affiliation{CNR-SPIN Institute for Superconductors, Innovative Materials and Devices, Corso Perrone 24, Genova, Italy}
\thanks{These two authors contributed equally}

\author{Giordano~\surname{Mattoni}$^{\dag}$}
\affiliation{Kavli Institute of Nanoscience, Delft University of Technology, P.O. Box 5046, 2600 GA Delft, The Netherlands}
\affiliation{Department of Physics, Graduate School of Science, Kyoto University, Kyoto 606-8502, Japan}

\author{Marco~\surname{Pelassa}}
\affiliation{Dipartimento Architettura e Design, Università degli Studi di Genova, Stradone S. Agostino 37, Genova, Italy}

\author{Warner~J.~\surname{Venstra}}
\affiliation{Kavli Institute of Nanoscience, Delft University of Technology, P.O. Box 5046, 2600 GA Delft, The Netherlands}
\affiliation{Quantified Air BV,  Rijnsburgersingel 77, 2316 XX Leiden, The Netherlands}

\author{Herre~S.~J.~\surname{van~der~Zant}}
\affiliation{Kavli Institute of Nanoscience, Delft University of Technology, P.O. Box 5046, 2600 GA Delft, The Netherlands}

\author{Andrea~D.~\surname{Caviglia}}
\affiliation{Kavli Institute of Nanoscience, Delft University of Technology, P.O. Box 5046, 2600 GA Delft, The Netherlands}


\begin{abstract}
  \textbf{This document is the Accepted Manuscript version of a
    Published Work that appeared in final form in ACS Applied
    Materials \& Interfaces, copyright \copyright~American Chemical
    Society after peer review and technical editing by the
    publisher. To access the final edited and published work see
    \href{https://pubs.acs.org/doi/10.1021/acsami.9b14501}{https://pubs.acs.org/doi/10.1021/acsami.9b14501}}.\newline\newline
  Strain engineering is one of the most effective approaches to
  manipulate the physical state of materials, control their electronic
  properties, and enable crucial functionalities.  Because of their
  rich phase diagrams arising from competing ground states, quantum
  materials are an ideal playground for on-demand material control,
  and can be used to develop emergent technologies, such as adaptive
  electronics or neuromorphic computing.  It was recently suggested
  that complex oxides could bring unprecedented functionalities to the
  field of nanomechanics, but the possibility of precisely controlling
  the stress state of materials is so far lacking.  Here we
  demonstrate the wide and reversible manipulation of the stress state
  of single-crystal \ce{WO3} by strain engineering controlled by
  catalytic hydrogenation.  Progressive incorporation of hydrogen in
  freestanding ultra-thin structures determines large variations of
  their mechanical resonance frequencies and induces static
  deformation.  Our results demonstrate hydrogen doping as a new
  paradigm to reversibly manipulate the mechanical properties of
  nanodevices based on materials control.
\end{abstract}

\maketitle

\section*{Introduction}
Complex oxides are characterised by a rich energy landscape governed
by multiple thermodynamic parameters, including temperature, stress,
chemical potential, and electromagnetic
fields.\cite{MacManus-Driscoll2008, Rondinelli2012, Ramesh2019} Phase
competition in these quantum materials leads to giant responses to
external stimuli associated with large and non-linear
susceptibilities.  Chemical doping is a powerful control parameter to
switch between their competing phases, and oxygen vacancies have often
been employed to induce changes in electrical, structural or magnetic
properties, although with limited reversibility.\cite{Manca2015,
  Swallow2017, Yao2017, Zhang2018} Hydrogen intercalation is an
alternative route to effectively control the ground state of these
materials, as an example, by stabilizing metallic phases and promoting
lattice symmetry,\cite{Kilic2002, Wei2012, Yoon2016} but the
possibility of precisely controlling the stress state is so far
lacking.  A particularly interesting system with multiple competing
ground states regulated by anharmonic couplings between different
structural distortions is \ce{WO3}.\cite{Hamdi2016} Its complex energy
landscape determines large changes of its lattice and electronic
properties in response to chemical doping,\cite{Mattoni2018a} electric
fields,\cite{Leng2017} and epitaxial strain.\cite{Du2014} These
characteristics have important technological applications, such as
electrochromic devices, smart windows,\cite{Svensson1984,
  Svensson1985, Granqvist2000} and gas sensing where record-holding
pmm sensitivity to \ce{H2} was recently
demonstrated.\cite{Mattoni2018b}

Here we show that the electromechanical response of freestanding
single-crystal \ce{WO3} microbridges can be reversibly controlled by
hydrogen gas at room temperature.  The incorporation of hydrogen in
\ce{WO3} thin films induces a change in the out-of-plane lattice
constant up to 1.3\%, an effect that we use to tune the mechanical
resonances of \ce{WO3} microbridges as their stress state changes from
tensile to compressive.

\section*{Results and Discussion}
This experiment is performed on a 50\,nm-thick single crystal \ce{WO3}
film grown on top of a Ti-terminated \ce{SrTiO3} (001) substrate.  The
\ce{WO3} thin film shows a flat surface with a step-and-terrace
morphology that indicates good heteroepitaxial growth (film
characterisation in Supporting Information, Sec.~I).  A small amount
of Pt, of equivalent thickness 0.2~nm, is then deposited by e-beam
evaporation to enable hydrogen intercalation, as described in the
Methods section and in Ref.~\citenum{Mattoni2018b}.  We note that the
hydrogen incorporation rate can be regulated by the amount of Pt
catalyst: here we choose to use a low amount in order to slow down the
process and monitor it as a function of time.  The single-crystal
character of \ce{WO3} is confirmed by X-ray characterisation, where
narrow rocking curves and reciprocal space maps show that the film is
coherently strained to the substrate lattice. $\theta$--$2\theta$
scans along the (001) and (002) peaks of \ce{WO3} present finite size
oscillations, indicating high crystal quality (see Supporting
Information, Sec.~II).  From the X-ray data we extract a $c$-axis
length of 3.70\,\AA, in agreement with previous reports of \ce{WO3}
thin films on \ce{SrTiO3}.\cite{Mattoni2018a} Considering that in its
bulk pseudocubic phase \ce{WO3} has a lattice constant of about
3.77\,\AA,\cite{Crichton2003} epitaxial lock imposed by the substrate
determines an elongation of the $a$ and $b$ axes of about 3.4\,\%,
with a consequent decrease of the $c$-axis that in our film amounts to
$-2$\,\%. The film is hence under in-plane tensile stress.  This
analysis indicates that the unit cell volume in thin films is slightly
larger than in bulk. This is expected and could be related to two
mechanism: a Poisson's ratio lower than 0.5 and the presence of oxygen
vacancies. While the former contribution is present in the vast
majority of compounds, the presence of oxygen vacancies is almost
unavoidable in \ce{WO3} thin films with optimal structural
quality,\cite{Mattoni2018a} and most probably constitute one of the
main relaxation mechanism for epitaxial strain.

\begin{figure}[]
  \includegraphics[width=86mm]{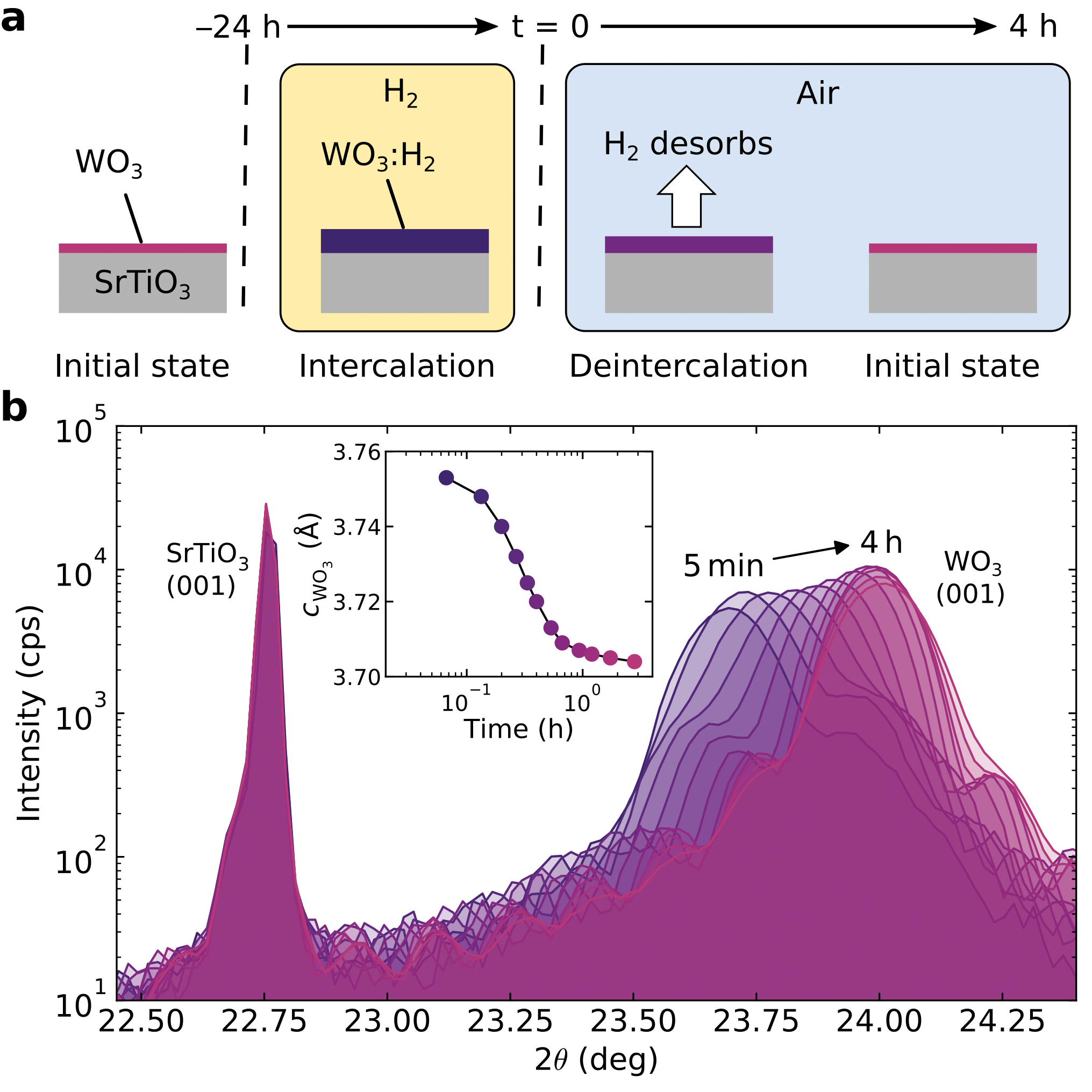}
  \caption{
    \label{fig:XRD}
    \textbf{Lattice expansion due to hydrogen doping in \ce{WO3} thin
      films.}  (a) Schematic representation of the experiment showing
    hydrogen intercalation in a \SI{20}{\percent} \ce{H2 / Ar} mixture
    and deintercalation in air.  (b) X-ray scan showing the sharp
    (001) peak of the \ce{SrTiO3} substrate and \ce{WO3(001)} peak
    with its finite size oscillations during hydrogen
    deintercalation. The inset shows the variation of \ce{WO3}
    $c$-axis length as a function of time.  }
\end{figure}

We measure the structural response of \ce{WO3} to \ce{H2} gas by
monitoring changes in its $c$-axis parameter by X-ray diffraction.
The experimental procedure comprises the two steps illustrated in
Figure~\ref{fig:XRD}(a): The sample is initially placed at room
temperature in a sealed chamber filled with 1\,bar of a 20\,\%
\ce{H2}/\ce{Ar} mixture for \SI{24}{\hour} (intercalation) to achieve
a stable (saturated) hydrogen doping condition.  When transferred to
the XRD setup, the sample is exposed to air at $t=0$.  In this step,
hydrogen is released from the \ce{WO3} lattice (deintercalation) and
the material progressively regains its initial state.
Fig.~\ref{fig:XRD}(b) shows several $\theta$--$2\theta$ scans taken
during hydrogen deintercalation, where a progressive shift of the
\ce{WO3(001)} peak and its finite size oscillations occurs.  This
shift corresponds to a change in \ce{WO3} $c$-axis parameter which is
reported as a function of time in Fig.~\ref{fig:XRD}(b).  In the
hydrogen-doped state at $t=5$\,min we measure $c_{WO_3}=3.75$\,\AA,
indicating an increase of about $1.25$\,\% with respect to the undoped
condition.  Upon hydrogen deintercalation, $c_{WO_3}$ progressively
decreases and recovers the value of the initial state after about
1\,h.  This indicates that hydrogen determines a large and reversible
expansion of the \ce{WO3} lattice, with a magnitude comparable to what
has been previously reported upon the formation of oxygen vacancies or
intercalation of alkali
metals.\cite{Crichton2003,Dass2013,Mattoni2018a}

\begin{figure*}[t]
  \includegraphics[width=172mm]{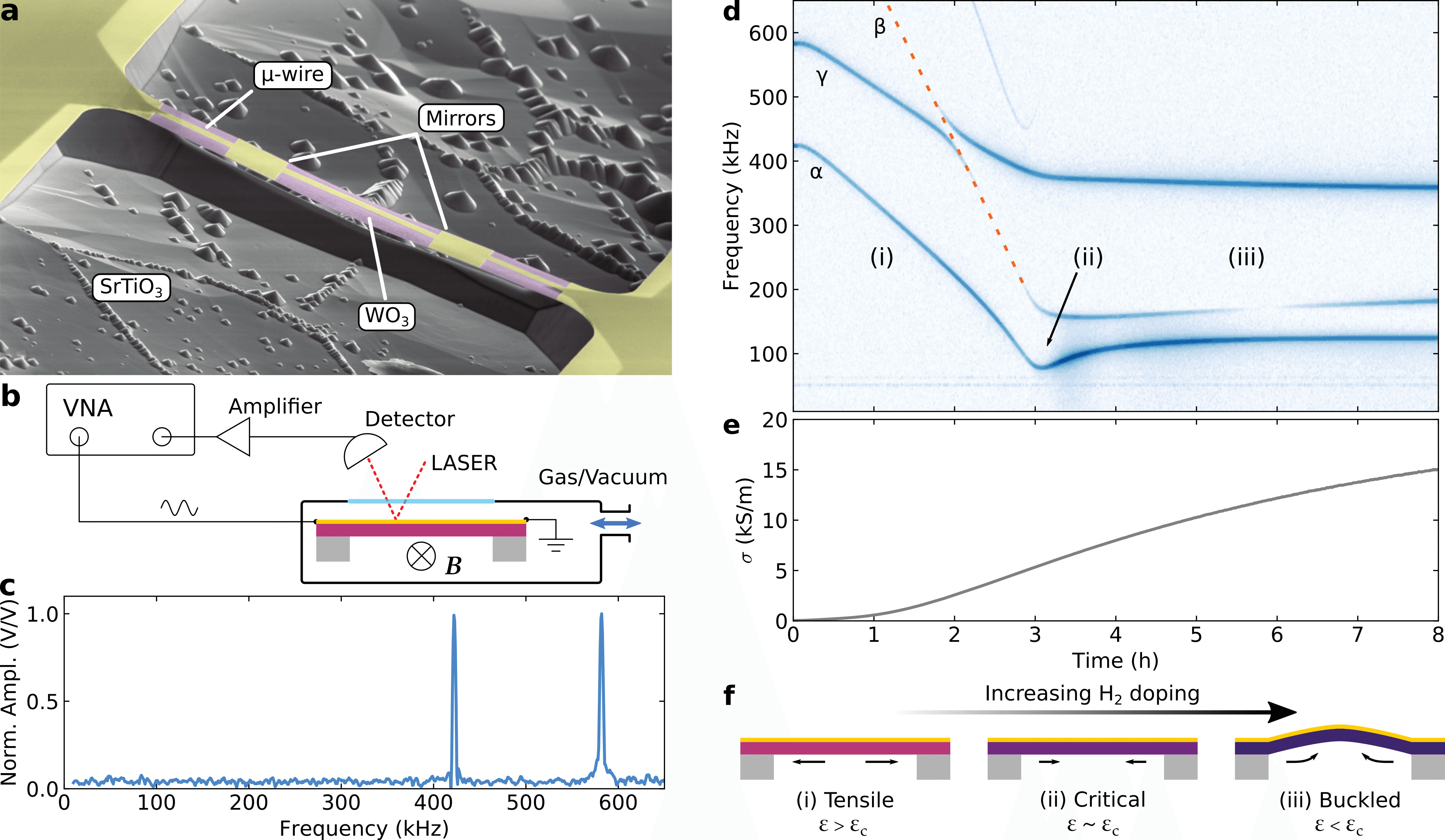}
  \caption{
    \label{fig:device}
    \textbf{Hydrogen doping of single-crystal \ce{WO3} microbridges.}
    (a) Scanning electron micrograph in false colors of the measured
    device.  (b) Schematic of the experimental setup (VNA, vector
    network analyser).  (c) Mechanical spectrum of a 110\,\textmu
    m-long microbridge in the pristine state.  (d) Colormap of the
    mechanical spectra of the microbridge in (a) as a function of time
    when exposed to 5\,mbar of \SI{20}{\percent} \ce{H2}/Ar gas
    mixture and (e) electrical conductivity measured at the same time.
    (f) Illustration of the three regimes identified in (d). The
    color-scale in (d) is linear and normalized in V/V.  }
\end{figure*}

The large lattice expansion induced by hydrogen doping offers the
interesting possibility to control the mechanical stress of the
material in a reversible manner.  To explore this possibility, we
realize \ce{WO3} freestanding microbridges.  The structures are
fabricated by using lithographic patterning and combined dry/wet
etching processes in order to remove the \ce{SrTiO3} substrate, a
procedure similar to the one used in previous reports.\cite{Manca2015}
(see also Methods section and Supporting Information, Sec.~III).
Figure~\ref{fig:device}(a) shows the false-colour micrograph of a
typical \ce{WO3} microbridge acquired by scanning electron microscopy.
The freestanding structure (5\,\textmu m-wide, 110\,\textmu m-long) is
composed of a 50\,nm-thick \ce{WO3} crystalline thin film (purple),
and 50\,nm-thick gold elements (yellow) comprising two
\SI{5x5}{\micro\metre} mirrors and a 1\,\textmu m-wide microwire that
runs throughout the freestanding region.  The mirrors are used to
reflect the laser light employed to measure the mechanical motion of
the bridge in an optical lever geometry (Fig.~\ref{fig:device}(b)),
while the microwire provides a low-impedance electrical channel
required for the electrical excitation.  By applying an alternating
electrical current through the gold wire, the microstructure is
mechanically actuated via the magnetomotive and electrothermal
mechanisms (see Methods).  The typical mechanical spectra of a
pristine 110\,\textmu m-long microbridge is shown in
Fig.~\ref{fig:device}(c).  We also fabricated microbridges of other
lengths between 50 and 110\,\textmu m, all of which show similar
spectra (see Supporting Information, Sec.~IV).  The fundamental
resonance mode of the longer beams shows a better signal-to-noise
ratio compared to the shorter ones, as expected due to a larger
deflection. Furthermore, a higher quality factor is observed, which is
attributed to lower clamping losses (see Supporting Information,
Sec.~V).  For this reason, we focus on the 110 \textmu m-long
microbridge to investigate the changes induced by hydrogen
incorporation.

The simultaneous mechanical and electrical characterization is
performed in a vacuum chamber with variable gas environment, optical
access, and a controlled sample temperature fixed at 25$^{\circ}$C.
The \ce{WO3} device is initially in an undoped condition and we
measure its mechanical spectrum as a function of time in
Figure~\ref{fig:device}(d). Details of the optical setup employed in
this experiment are reported in the Supporting Information, Sec~VI.
At $t=0$ we introduce a 20\,\%~\ce{H2}/Ar mixture at low pressure
(0.5\,mbar) to follow the hydrogen intercalation dynamics with minimal
damping of the mechanical motion.  Hydrogen doping results in a
dramatic change in the mechanical behaviour of the microbridge which
can be divided into three distinct regions: (i) a steep decrease of
the eigenfrequencies, (ii) a transitional regime, and (iii) a flat
response.  The lowest mechanical mode $\alpha$ has a smooth behaviour
across the different regions and corresponds to the first flexural
mode, as discussed in the following.  At $t=0$ the \ce{WO3}
microbridge is in a tensile strain state originating from the lattice
mismatch with the \ce{SrTiO3} substrate.  In this condition, the
structure can be schematically modelled as an ideal thin and long
double-clamped beam where the flexural resonance frequencies are given
by the Euler--Bernoulli equation\cite{Tilmans1992}
\begin{equation}
  \label{eq:strain_frequency}
  f_n(\varepsilon) =
  a_n \frac{t}{l^2}
  \sqrt{
    \frac{E}{\rho}
    \left(
    \frac{1}{(1-\nu^2)} +
    b_n \varepsilon \left(\frac{l}{t}\right)^2
    \right)
  },
\end{equation}
with resonance frequency $f_n$ relative to the $n^{th}$ mode, axial
strain $\varepsilon$, Young modulus $E$, density $\rho$, Poisson ratio
$\nu$, length $l$, and thickness $t$. $a_n$ and $b_n$ are numerical
coefficients related to the mode shape.
Eq.~(\ref{eq:strain_frequency}) shows that the mechanical
eigenfrequencies of the microbridge change as a function of strain
$\varepsilon$.  In our experiment $\varepsilon$ changes continuously
over time due to the large \ce{WO3} lattice expansion during hydrogen
intercalation (Fig.~\ref{fig:XRD}).  The most dramatic effect is
observed in region (i), where the relaxation of the initial tensile
strain results in a large drop of the resonance frequencies.
According to Eq.~(\ref{eq:strain_frequency}) the frequency of the
first flexural mode $f_1(\varepsilon)$ is expected to reach zero at
the critical strain value
\begin{equation}
  \label{eq:frequency_minimum}
  \varepsilon_\mathrm{c} = -\frac{1}{1-\nu^2}\left(\frac{t}{l}\right)^2\frac{1}{b_1}.
\end{equation}
However, in real systems this is typically not observed because device
asymmetry causes the frequency to reach a smooth minimum at the finite
strain value $\varepsilon_{\mathrm{c}}$.\cite{Kim1986} In our
experiment this occurs in region (ii) at $t\sim\SI{3}{\hour}$, where
the resonance frequency of the lowest mode $\alpha$ shows a minimum at
$f=\SI{80}{\hertz}$, about one fifth of its initial value. A further
increase of the compressive strain causes out-of-plane buckling. This
occurs in region (iii), where Eq.~{\ref{eq:strain_frequency}} is not
applicable and the resonance frequency of $\alpha$ increases again as
the excess compressive strain energy is stored in the form of
out-of-plane deformation (buckling).  At the onset of the buckled
state, the frequencies of the odd mechanical modes are expected to
rise slightly,\cite{Kim1986, Bouwstra1991, Nayfeh1995} consistent with
the small frequency increase that $\alpha$ undergoes in region (ii).
Multiple higher-order modes are visible in the spectral map of
Fig.~\ref{fig:device}(d), with frequencies above $\alpha$.  Their
frequencies cannot be represented by a simple string resonator because
interfacial stress and non-uniform mass distribution play an important
role.  These resonances are probably related to torsional and flexural
modes of even order which show different response to
strain,\cite{Bouwstra1991} as discussed in the last part of this work.
An avoided crossing between modes $\beta$ and $\gamma$ occurs at
around $t=\SI{2}{\hour}$, indicating a strong coupling between the
mechanical modes. This feature arises as each mode shows a different
tuning slope with applied compressive stress.\cite{Bouwstra1991,
  Nayfeh1995, Lacarbonara2005} Finally, we note that the data reported
in Fig.~\ref{fig:device}(d) are in good qualitative agreement with
similar measurements performed during hydrogen deintercalation in air,
thus showing good reversibility of the process (Supporting
Information, Sec.~VII).

During hydrogen intercalation we also measure the time dependence of
the electrical conductivity of \ce{WO3} ($\sigma_{\ce{WO3}}$), which
is reported in Fig.~\ref{fig:device}(e).  In our device design, the
conductivity of the microbridge is always determined by the gold
microwire, whose constant electrical resistance (50~$\Omega$) is much
lower than that of \ce{WO3} at any level of hydrogen doping. For this
reason, we monitor $\sigma_{\ce{WO3}}$ on a separate region of the
film close to the microbridge.  The data shows the progressive
metallization of the material due to electronic doping, in agreement
with previous reports.\cite{Mattoni2018b} Hydrogen intercalation
progresses smoothly during the whole experiment and saturation occurs
only towards the end of the measured time span ($t > 8$\,h).  This
indicates that between $t=4$ and $8$\,h the amount of hydrogen in the
\ce{WO3} lattice is significantly increasing, even if the mechanical
modes show a flat response.  We note that, albeit the suspended
\ce{WO3} microbridge has a higher surface area exposed to \ce{H2} gas
compared to the clamped film, in both cases the \ce{Pt} catalyst is
present only on the top surface, thus determining the same rate of
hydrogen intercalation.  Furthermore, the rate of hydrogen dynamics in
this experiment is significantly longer than during the XRD
measurements of Fig.~\ref{fig:XRD} because of the low hydrogen
pressure, and the different rates of intercalation and
deintercalation.\cite{Mattoni2018b} The observed evolution of the
microbridge mechanical states is schematically summarised in
Fig.~\ref{fig:device}(f).

\begin{figure}[t]
  \includegraphics[width=86mm]{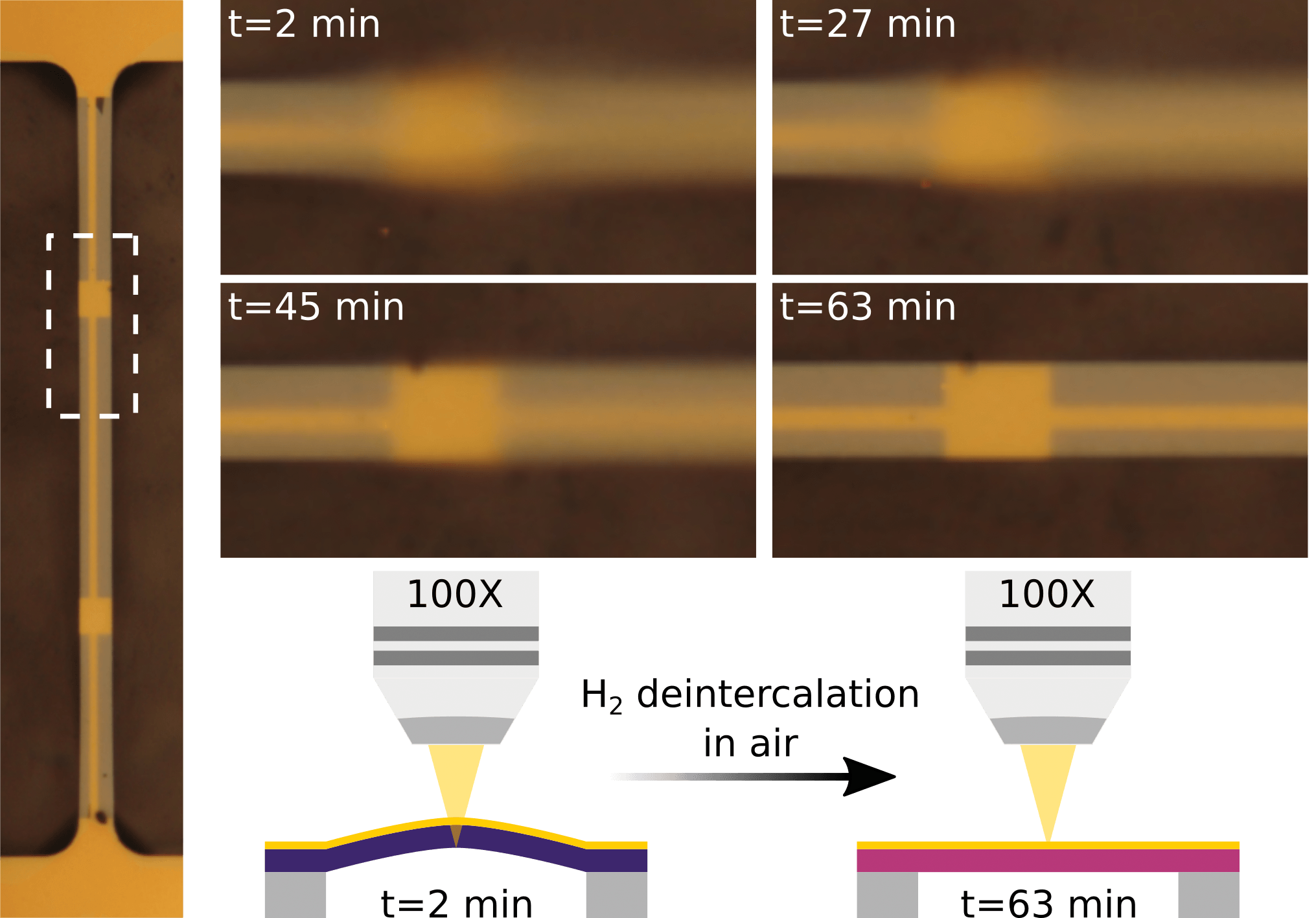}
  \caption{
    \label{fig:optical}
    \textbf{Relaxation of the buckled state upon hydrogen
      deintercalation.}  Optical images of the \ce{WO3} microbridge of
    Fig.~\ref{fig:device}(a) initially in a condition of saturated
    doping and exposed to air at $t=0$. The white dashed rectangle
    indicates the magnified region.  The schematic drawing illustrates
    how relaxation of the buckled state brings back the microbridge in
    the microscope focal plane.  }
\end{figure}

A strong evidence of the insurgence of a buckled state upon the
intercalation of hydrogen can be obtained by optical microscopy. For
this purpose, we report in Figure~\ref{fig:optical} a series of
photographs showing one gold mirror of the microbridge.  For this
experiment the microbridge is first prepared in a saturated-doping
state and at $t=0$ it is exposed to air at room temperature, the same
procedure used in Fig.~\ref{fig:XRD}.  The photographs are acquired at
different times with an objective lens (depth of field of about
\SI{0.7}{\micro\metre}) whose focal plane is fixed to have in focus
the clamped edges of the microbridge.  In the initial state at
$t=2$\,min the mirror is out of focus, indicating static deformation
in the direction perpendicular to the focal plane.  The structure
gradually gets more in focus at $t=45$\,min, and at $t=63$ min, after
about one hour in air, the whole microbridge is in focus, indicating
the recovery of a flat state in the undoped conditions.  We note that
this time span of hydrogen deintercalation in 1\,bar of air is
comparable to the one of the XRD measurements of Fig.~\ref{fig:XRD}
and, as previously discussed, much faster than the slow dynamics
observed in Fig.~\ref{fig:device}.

\begin{figure}[t]
  \includegraphics[width=86mm]{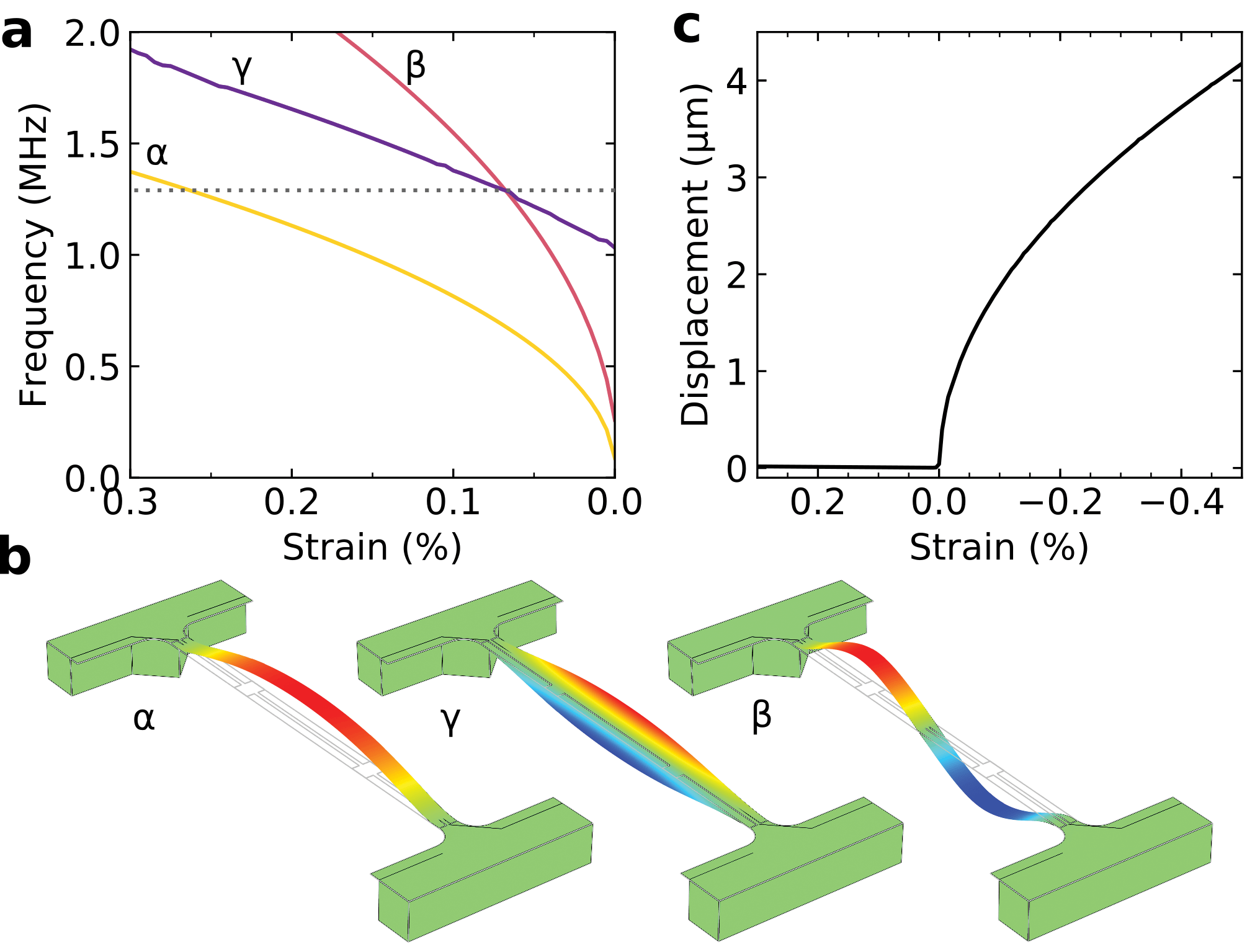}
  \caption{
    \label{fig:fea}
    \textbf{Finite element simulation of a strained \ce{WO3}
      microbridge.}  (a) Strain-dependence of the first flexural
    ($\alpha$, $\beta$) and torsional ($\gamma$) mechanical modes and
    (b) corresponding mode shapes.  (c) Vertical displacement of the
    microbridge center as a function of strain.  }
\end{figure}

To better understand the changes of the \ce{WO3} microbridge
mechanical properties, we perform a finite element analysis as a
function of strain (details in Methods and Supporting Information,
Sec.~VIII).  The strain-dependence of the three lowest modes of the
simulated device and the corresponding mode shapes are reported in
Figures~\ref{fig:fea}(a) and \ref{fig:fea}(b), respectively.  The
trend of the frequencies well represents our experimental data, with a
rapid decrease of the first and second flexural modes $\alpha$,
$\beta$, and a slower variation of the torsional mode $\gamma$.  Due
to their different slope, $\beta$ and $\gamma$ intersect at about
$f=1.3$\,MHz (intersection of mode $\alpha$ with the grey dotted
line).  Also the presence of a mode crossing is consistent with our
experimental results, where it occurs around $f=0.45$\,MHz
(Fig.~\ref{fig:device}(d)).  We note that the experimental crossing
frequency is about three times smaller than the simulated one, a
discrepancy probably caused by the fact that we used values of
$E_{WO_3}$ and $\nu_{WO_3}$ from first principles since no
experimental data is available in literature.\cite{Liu2018} However,
the strain value corresponding to the crossing point in our
simulations is weakly dependent on $E$ and $\nu$ (Supporting
Information, Sec~VIII), and can thus be used as a reference to
estimate the in-plane strain of our experimental device. As an
additional reference point, we note that in the experimental data of
Fig.~\ref{fig:device}(d) the frequency of mode $\alpha$ at $t=0$
corresponds approximately to the frequency of the avoided mode
crossing. Using this information in Figure~\ref{fig:fea}(a), we can
estimate the initial strain of the microbridge to be about
$\varepsilon_0=0.25$\,\% (grey dotted line). Surprisingly, this value
is rather small compared to the epitaxial mismatch of 3.4\,\% between
\ce{WO3} and the \ce{SrTiO3} substrate.  However, this can be
explained considering that oxide compounds can easily accommodate
stress with the formation of dislocations or point defects, such as
oxygen vacancies.  In particular, previous work shows that a low
concentration of oxygen vacancies during the growth of \ce{WO3} is
sufficient to determine a relatively large lattice expansion, thus
effectively relaxing the strain while maintaining a high crystal
quality.\cite{Mattoni2018a}

It is possible to estimate the total experimental variation of the
microbridge strain from the $c$-axis expansion $\Delta c$ measured by
XRD (Fig.~\ref{fig:XRD}).  For this purpose, we note that by taking
into account the different constraints for an epitaxial film and a
freestanding structure we have
\begin{equation}
  \label{eq:strain_c}
  \frac{\Delta c}{c_0} = \left( \frac{1+\nu}{1-\nu} \right) \Delta \varepsilon,
\end{equation}
with out-of-plane lattice parameter in the undoped state $c_0$,
Poisson ratio $\nu$, and total strain variation $\Delta \varepsilon$
(see Supporting Information, Sec.~IX).  With the parameters employed
in the simulation we obtain $\Delta \varepsilon = 0.75\,\%$.  Taking
the initial microbridge strain as +0.25\,\% for the discussion above,
we can estimate the strain in the saturated doping condition to be
about \SI{-0.5}{\percent}.  On the basis of this analysis, we
calculate the vertical displacement of the microbridge centre as a
function of strain in Fig.~\ref{fig:fea}(c). For positive (tensile)
strain the displacement is negligible, while beyond the critical value
$\varepsilon_\mathrm{c}$ the structure relaxes the negative
(compressive) strain by bending.  At $\varepsilon=-0.5$\,\% the
simulation gives a displacement of 4\,\textmu m, a value compatible
with the optical measurements of Fig.~\ref{fig:optical}.  We note that
the crossing point between tensile and compressive strain
($\varepsilon_\mathrm{c} \sim 0$) lies at one third of the estimated
range of experimental strain in Fig.~\ref{fig:fea}(c). This is in good
agreement with the experimental data of Fig.~\ref{fig:device}(d),
where the critical condition of region (ii) lies also at about one
third of both the total time and the electrical conductivity spans.
Finally, we emphasize that in MEMS/NEMS devices it is particularly
difficult to generate large compressive stress, a condition that
usually requires additional components such as electrostatic
actuators, piezoelectric elements or resistive heaters.  These
components present important drawbacks such as difficulty of focusing
the actuation force on the resonator beam, hurdles in high-purity
material fabrication, and insurgence of high temperatures.  Our
approach involving chemical doping proved to be a fundamental step
forward over the present difficulties, providing a pathway for strong
and localized strain-control of micro and nanodevices.

\section*{Conclusion}
In summary, we demonstrated a new effective approach for controlling
in situ the stress state of oxide-based freestanding structures.  We
achieved reversible control of the mechanical properties of
single-crystal \ce{WO3} microbridges, where the large lattice
expansion due to hydrogen intercalation allowed us to finely tune the
stress state from tensile to compressive, with a final buckled
configuration.  The proposed approach could be extended in two
directions. By exploring the mechanical response of other oxides
micromechanical systems to hydrogen doping. By realizing full-oxide
heterostructures comprising a \ce{WO3} layer acting as functional
strain-tuning element.  Our work highlights the potential of complex
oxides to realise tunable nanomechanical systems or novel sensing
devices.

\section*{Methods}

\paragraph*{Device fabrication:}

The single-crystal 50\,nm-thick \ce{WO3} thin film on
\ce{TiO2}-terminated \ce{SrTiO3}(001) substrate was grown by pulsed
laser deposition using \SI{500}{\celsius} substrate temperature,
\SI{8e-2}{\milli\bar} oxygen pressure,
\SI{1}{\joule\per\square\centi\metre} laser fluence, and
\SI{1}{\hertz} repetition rate.  Pt was deposited by e-beam
evaporation, its growth rate was monitored using a quartz
microbalance. The process was stopped at an equivalent thickness of
0.2~nm.  The microbridge structures were realised by a two-step
electron-beam patterning using PMMA resist.  The first patterning step
was used to define the geometry of high-reflectivity mirrors and
low-impedance microwire, followed by electron-beam evaporation of a
\SI{5/45}{\nano\metre} \ce{Ti/Au} overlayer and lift-off in acetone.
The second patterning step defined the microbridge mesa and was
followed by a 6\,min \ce{Ar} etching process (10\,mA beam current,
500\,eV beam energy, and $7\times10^{-4}$\,mbar pure Ar atmosphere),
during which a total of 60\,nm of \ce{WO3/SrTiO3} heterostructure was
removed.  The microbridges are finally made freestanding by
selectively etching the \ce{SrTiO3} substrate in a \SI{4}{\percent}
aqueous solution of \ce{HF} at \SI{30}{\celsius} for \SI{30}{\minute},
resulting in a total of \SI{7}{\micro\metre} vertical distance between
the etched substrate and the freestanding \ce{WO3}.

\paragraph*{Measurement setup:}

The electrical conductance measurements were performed using a
four-probe DC configuration on a patterned square of the clamped
\ce{WO3} film.  Contacts to the \ce{WO3} film were provided by
wire-bonding \ce{Ti / Au} metal pads.  The mechanical measurements
were performed in a custom setup featuring a controlled gas
atmosphere, temperature stability of $\pm$50\,mK, motion detection of
microstructures with optical lever technique and electrical
pass-through.  The optical images of Fig.~\ref{fig:optical} were
acquired with an Olympus microscope equipped with an UMPlan~FI
100X/0.90 objective lens.

\paragraph*{Mechanical measurements:}

The mechanical modes of the microbridges are excited by biasing the
device with an AC voltage across the metal microwire of the form
$V(t)=V_0(1+\sin(\omega t)$).  Two concurrent mechanisms determine the
appearance of a mechanical force: magnetomotive and electro-thermal.
The magnetomotive mechanism arises from the Lorentz force exerted on
the current flowing in the microbridge by the magnetic field of a
permanent magnet located in the sample chamber.  The electrothermal
mechanism is due to the periodic thermal expansion and contraction of
the microbridge determined by periodic Joule heating.  Since the
resistance of the metal microwire ($\sim 50\,\Omega$) much lower than
the one of the \ce{WO3} bridge, even in its most-doped state, the
current flowing in the wire does not depend on \ce{WO3} doping
conditions, hence providing a constant excitation magnitude.  We note
that the film thickness of 50~nm was chosen to have resonance
frequencies falling within the accessible measurement range of our
detector (2~MHz) and homogenous doping in the out-pf-plane
direction. Thinner samples would improve both aspects, but with the
risk of lowering the fabrication yield.  The data in
Fig.~\ref{fig:device} were acquired using a HP4395a vector network
analyser with a 1\,kHz bandwidth and 5 averages taken over a period of
\SI{20}{\second}. Details of the optical setup are reported in the
Supporting Information, Sec.~VI.

\paragraph*{Finite element simulation:}

Finite element analysis was performed in Comsol
Multiphysics\textregistered~using the structural mechanic module. The
calculations employed the ``static'' and ``prestressed
eigenfrequency'' solvers with a parametric sweep of the strain
value. The mechanical parameters used to model the \ce{WO3}
microbridge are $E=\SI{300}{\giga\pascal}$, $\nu=0.25$ and
$\rho = \SI{7600}{\kilo\gram\per\cubic\meter}$, while Au was modelled
using the standard material library provided by the software. Further
details and device geometry are discussed in the Supporting
Information, Sec.~VIII.

\section*{Contributions}

N.M. and G.M. designed the experiment and performed the mechanical
measurements. N.M. designed the device geometry and performed the
finite element analysis. G.M. fabricated the device and performed the
XRD, SEM, and optical measurements. M.P. provided the structural
analysis leading to Eq.~(\ref{eq:strain_c}). W.J.V. designed and
realized the mechanical characterization setup. N.M. and G.M. wrote
the manuscript, with inputs from all authors.

\section*{Acknowledgement}
We thank P.~G.~Steeneken, L.~Pellegrino, and D.~Marré for helpful
discussions and valuable comments on the manuscript.  This work was
supported by The Netherlands Organisation for Scientific Research
(NWO/OCW) as part of the Frontiers of Nanoscience program
(NanoFront). This work was supported by the EU through the European
Research Council Advanced grant No. 677458 (AlterMateria). We
acknowledge received funding from the project Quantox of QuantERA
ERA-NET Cofund in Quantum Technologies (Grant Agreement N. 731473)
implemented within the EU H2020 Programme.

\section*{Supporting Infornation}
Growth and surface analysis of \ce{WO3} thin film, structural analysis
by X-ray diffraction, device fabrication, pictures of the final
devices, mechanical properties of microbridges of different lengths,
details of the finite element model, and analytic model of strain in
clamped and free-standing thin films.

\bibliography{library}

\newpage\newpage



\section*{Supplementary material}

# Supporting Information

—

# Large tunability of strain in $WO_3$ single-crystal microresonators controlled by exposure to $H_2$ gas

Nicola Manca[†,1,2,3,*] Giordano Mattoni[†,1,4] Marco Pelassa,[5] Warner J. Venstra,[1,6] Herre S. J. van der Zant,[1] and Andrea D. Caviglia[1]

[1]*Kavli Institute of Nanoscience, Delft University of Technology,*
*P.O. Box 5046, 2600 GA Delft, The Netherlands*
[2]*Dipartimento di Fisica, Università degli Studi di Genova, via Dodecaneso 33, Genova, Italy*
[3]*CNR-SPIN Institute for Superconductors,*
*Innovative Materials and Devices, Corso Perrone 24, Genova, Italy*
[4]*Department of Physics, Graduate School of Science,*
*Kyoto University, Kyoto 606-8502, Japan*
[5]*Dipartimento Architettura e Design, Università degli Studi di Genova,*
*Stradone S. Agostino 37, Genoa, Italy*
[6]*Quantified Air BV, Rijnsburgersingel 77, 2316 XX Leiden, The Netherlands*
(Dated: October 28, 2019)

---

[*] manca@fisica.unige.it;

[†] Authors contributed equally

This supplemental material contains the following:

**Supporting Information, Sec. I.  Growth and surface analysis of $WO_3$ thin film**

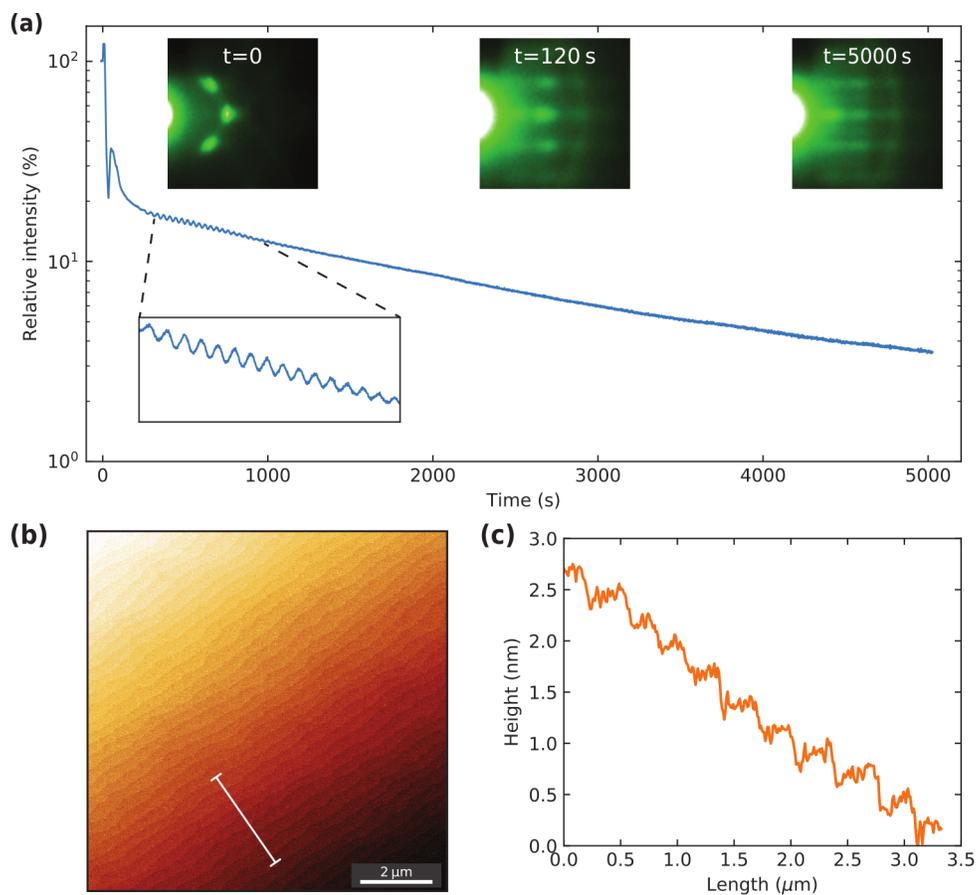

Figure S1. **Surface analysis of the $WO_3$ thin films**

Figure S1(a) shows the intensity over time of the RHEED (reflection high-energy electron diffraction) spot during the $WO_3$ growth. Intensity oscillations were employed to obtain a film with the desired thickness of about 50 nm, wich was later confirmed by X-ray diffraction. The RHEED pattern before ($t = 0\,\text{s}$), at the beginning ($t = 120\,\text{s}$), and at the end of the deposition ($t = 5000\,\text{s}$) is shown as insets. After the deposition, the sample surface was inspected by atomic force microscopy *ex situ*. Figure S1(b) shows the typical step-and-terrace structure of the $SrTiO_3$ substrate with single-unit-cell steps revealed by the line profile in Figure S1(c), indicating good crystal quality.

**Supporting Information, Sec. II.  Structural analysis by X-ray diffraction**

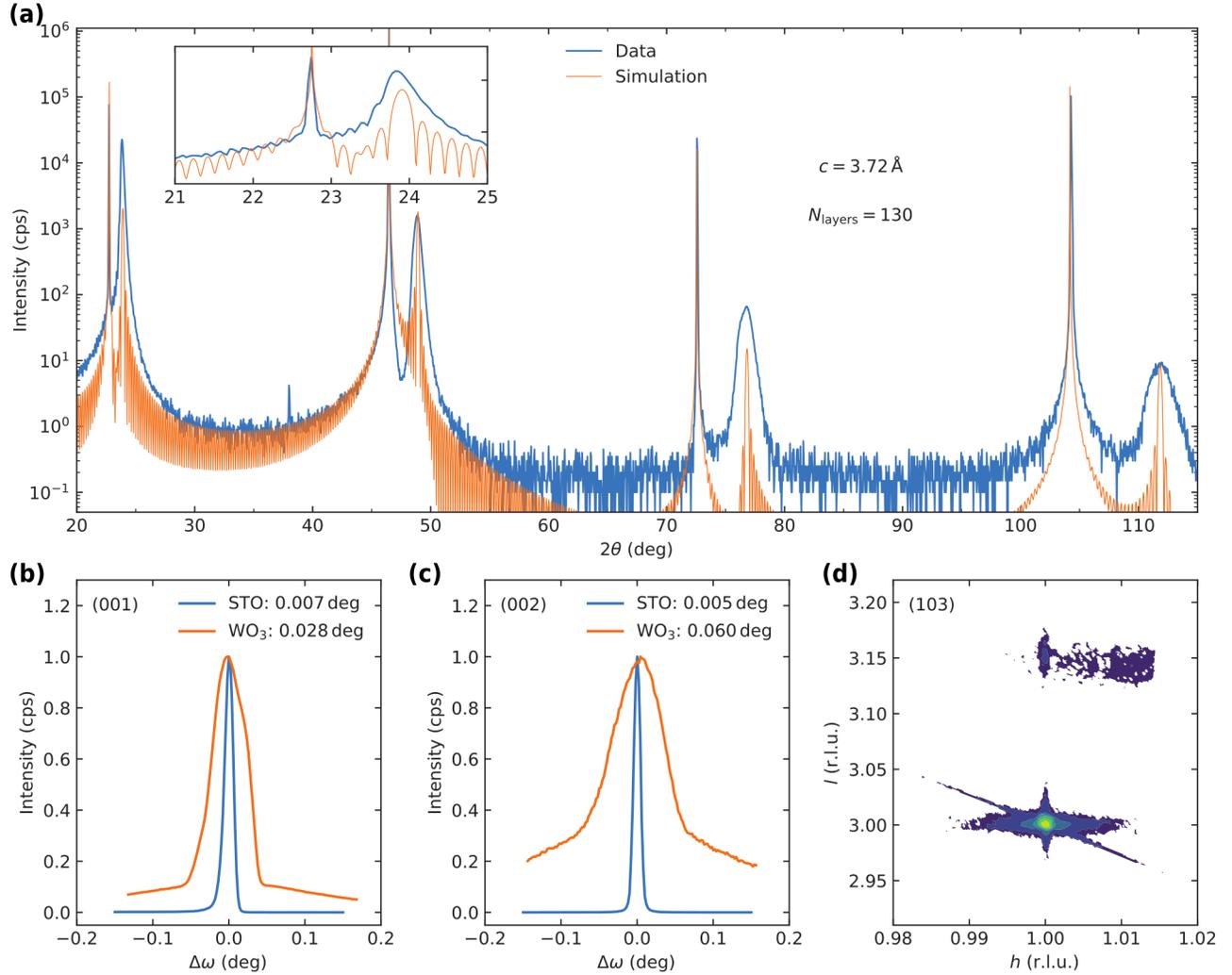

Figure S2. **X-ray diffraction measurements of the WO$_3$ thin film on SrTiO$_3$(001) susbtrate.**

In Figure S2 we report a structural characterization of the WO$_3$ film by X-ray diffraction. Figure S2(a) is a $\theta - 2\theta$ scan showing the first four diffraction peaks of the SrTiO$_3$ substrate and WO$_3$ thin film. We compare the experimental measurement (blue line) with a simulated data (orange line) calculated with a classical kinematic scattering model with pseudocubic structure factor for the SrTiO$_3$ substrate and for an N-layer-thick WO$_3$ film. The atomic scattering factors are calculated using the data from http://wwwisis2.isis.rl.ac.uk/reference/Xray_scatfac.htm. The model does not include strain relaxation and temperature effects. A magnification of the XRD data around the (001) peak is presented in the inset of Figure S2(a), where it is evident how the period

of the simulated finite size oscillations matches with the experimental data. Figure S2(b) and (e) show the rocking curves of the SrTiO$_3$ substrate and the WO$_3$ film around the (001) and (002) peaks, respectively, which have a remarkably low and comparable width. This indicates high coherence in the alignment of the lattice planes and it is a measure of single-crystal quality. The reciprocal space map in Figure S2(d) around the (103) diffraction peak of the substrate shows that the WO$_3$ peak is aligned along the same in-plane axis, indicating coherence between the substrate and film lattices.

**Supporting Information, Sec. III.   Device fabrication**

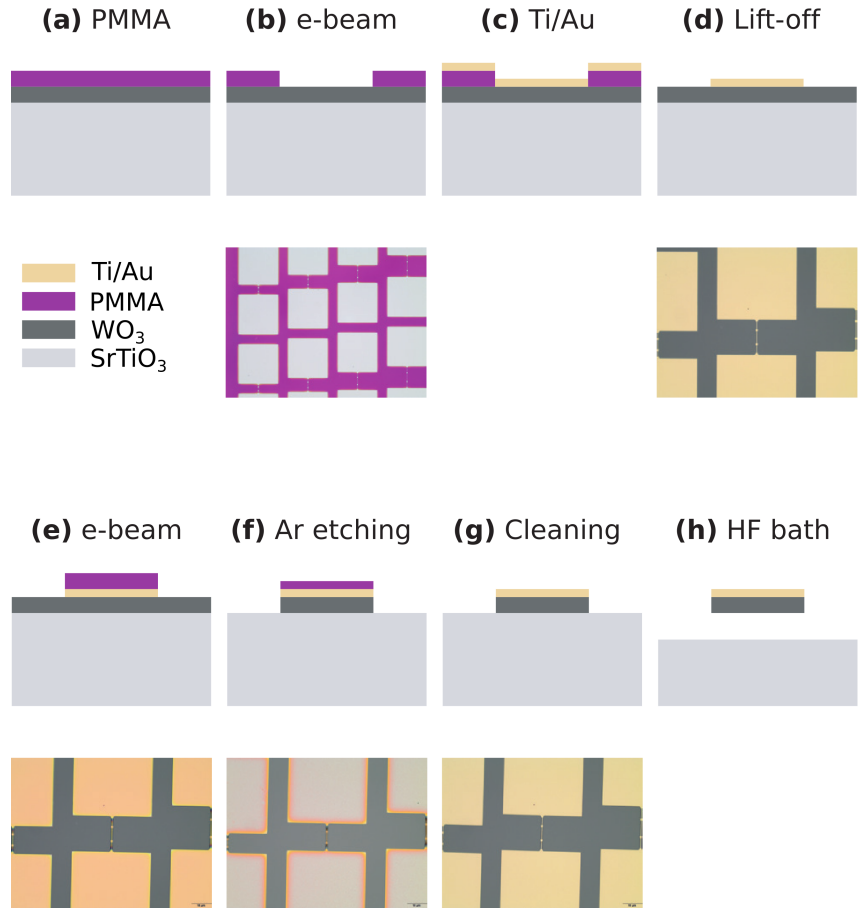

Figure S3.  **Fabrication process of the $WO_3$ suspended microbridges.** Schematic drawing of the main fabrication steps discussed in this section (top) and corresponding optical images of the actual sample (bottom). The optical colors depend on the exposure conditions and differ across the pictures.

Figure S3 shows the main steps for the fabrication of the $WO_3$ suspended microbridges. Starting from a 50 nm single-crystal thin film of $WO_3$ grown on top of a $SrTiO_3$ (001) substrate, we first deposit 0.2 nm of Pt by thermal evaporation. This small amount does not realize a connected layer but provide Pt centres that catalyse the $H_2$ dissociation. Fabrication steps proceed as follows:

(a) we deposit a layer of PMMA (A6-495) using a spin coater with revolution speed of 6000 rpm for about 50 s, which is baked at 180 °C for 10 min. Then, a thin layer of

*Elektra 92* is deposited on top of the PMMA by spin coating, followed by a baking for 60 s at 100 °C.

(b) the pattern for metallic regions is realized by electron beam (e-beam) lithography with a dose of 850 µC/cm$^2$.

(c) a metal layer of 5 nm Ti and 45 nm Au is deposited by e-beam evaporation.

(d) lift-off in warm acetone for 20 min at 60 °C.

(e) a second e-beam step defines the microbridges pattern which is aligned with the metal layer.

(f) to define the microbridge mesa we employed a dry etching technique (Ar milling, 500 eV, 0.2 mA/cm$^2$).

(g) a final cleaning was performed with NMP at 75 °C for 30min, followed by 10 min sonication in a ultrasound bath.

(h) the structures are made suspended by soaking in an acid bath of HF (4 % in water) at 30 °C for 30 min with a magnetic stirrer spinning at 100 rpm. HF selectovely etches the SrTiO$_3$ substrate without affecting the WO$_3$ film. Finally, the sample is dried using a critical point dryer.

**Supporting Information, Sec. IV.   Final devices**

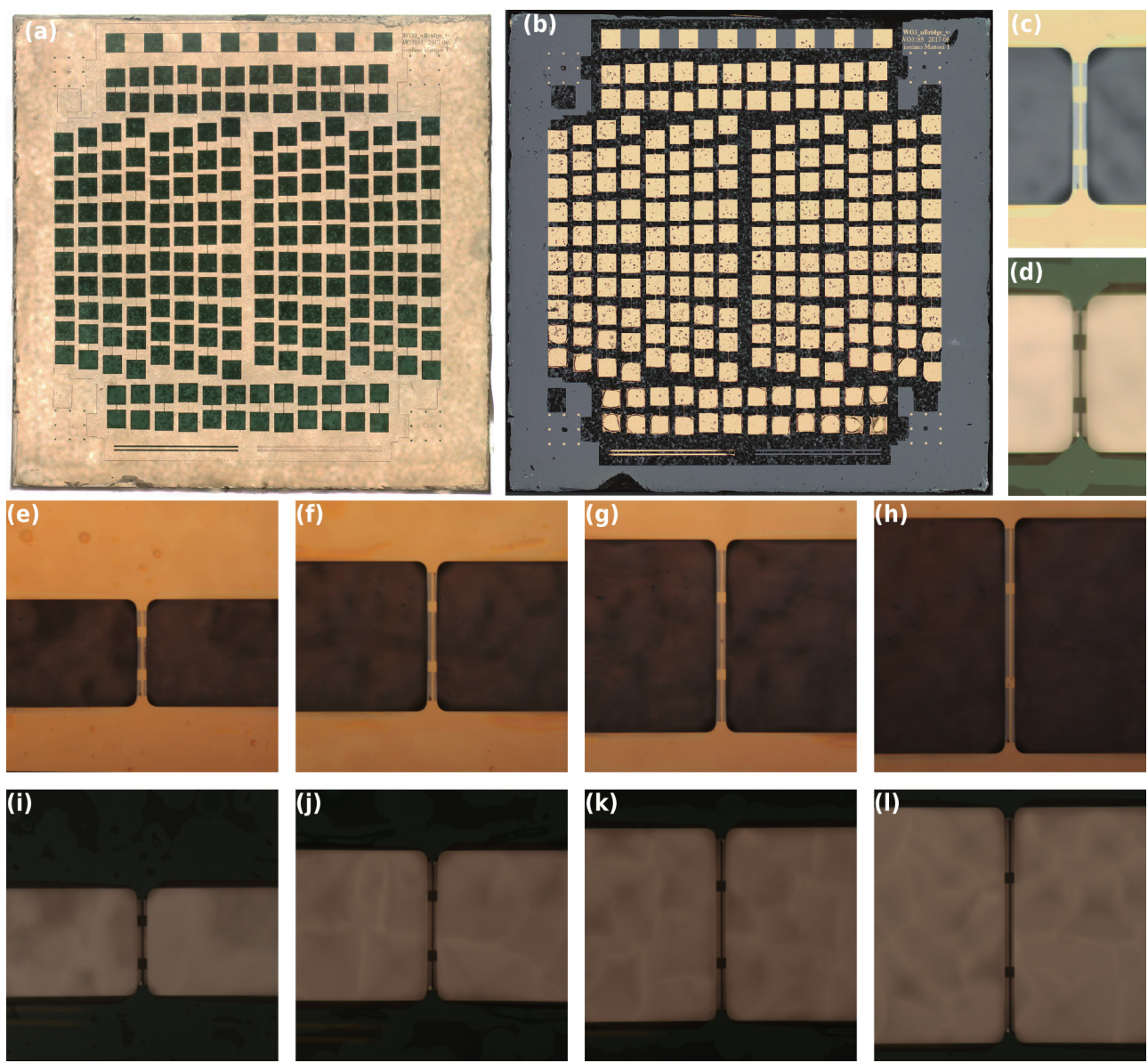

Figure S4.  **Sample after the fabrication process** Pictures of the 5×5 mm$^2$ sample in (a) transmitted and (b) reflected light. Pictures of a 50 µm-long microbridge acquired in (c) reflected and (d) transmitted light. (e–l) Pictures of microbridges having different length (50 µm, 70 µm, 90 µm, and 110 µm) acquired in reflected (top) and transmitted (bottom) light.


**Supporting Information, Sec. V.  Mechanical properties of microbridges for different length**

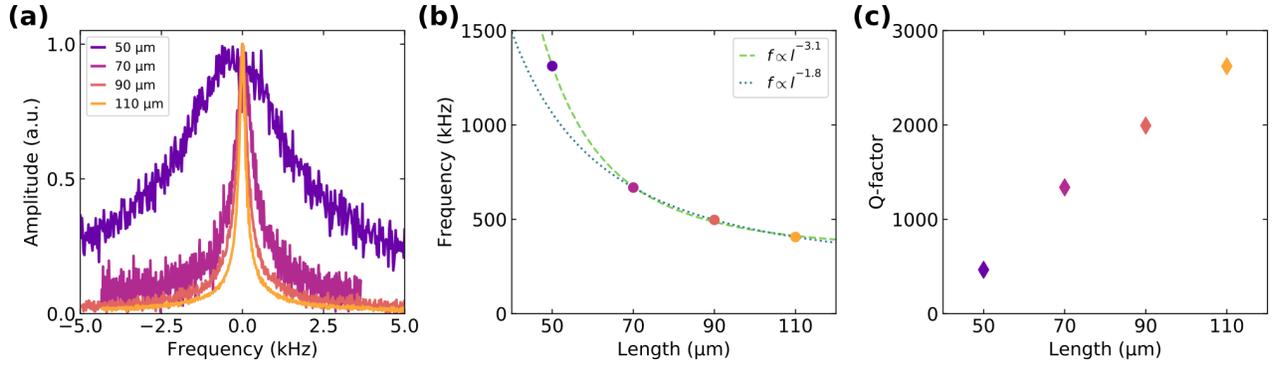

Figure S5. **First flexural mode for different bridge length** (a) Frequency response when driving beams of four different lengths near their fundamental flexural mode. The curves are normalized to the resonance frequency and the $WO_3$ material is in its pristine undoped state. (b) Extracted length dependence of the fundamental mode. The green dashed line is a fit with the power law $f \propto l^{-x}$ on the whole data range, while the blue dotted curve is a fit excluding the $l = 50\,\mu m$ microbridge (blue dotted curve). (c) Extracted Q-factor for the different devices.

In Figure S5 we investigate the length dependence of the resonance frequency and quality factor of our microbridges. Eq. (1) reported in the main text has two limiting cases with respect to $s$, leading to the power law dependence

$$f_n \propto 1/l^p$$

where $p$ equals 1 or 2 if $s$ is dominant ('string' limit) or negligible ('beam' limit), respectively. By fitting the data in Fig. S5(b) we find $p = 3.1$ (green dashed line). This value lays outside the expected range, indicating a deviation from the ideal behavior. The deviation from the expected behaviour could be the result of the non-uniform geometry due to the localized mass of the gold mirrors, or losses due to the clamping region. Furthermore, the triangular undercut of the $SrTiO_3$ close to the clamping may be partially involved in the beam motion. Finally, an initial curvature may be present as a result of incorporated stress. The origin of the stress could be thermal mismatch during fabrication, stresses at the $Au/WO_3$ interface, and/or non-symmetric strain relaxation through the thickness of the $WO_3$ beam. We tentatively fit the experimental data by excluding the shortest ($l$=50 µm) bridge and find $p = 1.8$ (blue dotted line). The fact that this value lays inside

the expected range supports the idea that non-ideal boundary conditions, which are more relevant for shorter structures, are a critical parameter affecting the mechanical characteristics of our devices. This observation is corroborated by the trend of the corresponding Q-factor reported in Fig. S5(c), that increases from about 450 for $l = 50\,\mu m$ up to 2600 for $l = 110\,\mu m$. Longer structures are thus not only intrinsically more sensitive to stress variations, as shown by Eq. (1) in the main text, but also have lower mechanical dissipation and thus are more suitable to measure changes induced by $H_2$ gas.



\end{document}